# Smart Guiding Glasses for Visually Impaired People in Indoor Environment

Jinqiang Bai, Shiguo Lian, *Member*, *IEEE*, Zhaoxiang Liu, Kai Wang, Dijun Liu

*Abstract*—To overcome the travelling difficulty for the visually impaired group, this paper presents a novel ETA (Electronic Travel Aids)-smart guiding device in the shape of a pair of eyeglasses for giving these people guidance efficiently and safely. Different from existing works, a novel multi-sensor fusion based obstacle avoiding algorithm is proposed, which utilizes both the depth sensor and ultrasonic sensor to solve the problems of detecting small obstacles, and transparent obstacles, e.g. the French door. For totally blind people, three kinds of auditory cues were developed to inform the direction where they can go ahead. Whereas for weak sighted people, visual enhancement which leverages the AR (Augment Reality) technique and integrates the traversable direction is adopted. The prototype consisting of a pair of display glasses and several low-cost sensors is developed, and its efficiency and accuracy were tested by a number of users. The experimental results show that the smart guiding glasses can effectively improve the user's travelling experience in complicated indoor environment. Thus it serves as a consumer device for helping the visually impaired people to travel safely.

*Index Terms*—AR, depth sensor, ETA, sensor fusion, vision enhancement

## I. INTRODUCTION

ACCORDING to the official statistics from World Health Organization, there are about 285 million visually impaired persons in the world up to the year of 2011: about 39 million are completely blind and 246 million have weak sight [1]. This number will increase rapidly as the baby boomer generation ages [2]. These visually impaired people have great difficulty in perceiving and interacting with the surroundings, especially those which are unfamiliar. Fortunately, there are some navigation systems or tools available for visually impaired individuals. Traditionally, most people rely on the white cane for local navigation, constantly swaying it in front for obstacle detection [3]. However, they cannot adequately perceive all the necessary information such as volume or distance, etc. [4]. Comparably, ETA (Electronic Travel Aid) can provide more information about the surroundings by integrating multiple electronic sensors and have proved to be effective on improving the visually impaired person's daily life [4], and the device presented in this work belongs to such category.

The RGB-D (Red, Green, Blue and Depth) sensor based ETA [5], [6] can detect obstacles more easily and precisely than other sensor (e.g. ultrasonic sensor, mono-camera, etc.) based scheme. However, a drawback of the depth sensor is that it has a limited working range in measuring the distance of the obstacle and cannot work well in the face of transparent objects, such as glass, French window, French door, etc. To overcome this limitation, a multi-sensor fusion based obstacle avoiding algorithm, which utilizes both the depth sensor and the ultrasonic sensor, is proposed in this work.

The totally blind people can be informed through auditory and/or tactile sensor [7]. Tactile feedback does not block the auditory sense, which is the most important perceptual input source. However, such an approach has the drawbacks of high power consumption and large size, which is not suitable for wearable ETA (like the glasses proposed in this work). Thus, sound or synthetic voice is the option for the first case. Some sound feedback based ETAs map the processed RGB image and/or depth image to acoustic patterns [8] or semantic speech [9] for helping the blind to perceive the surroundings. But, the blind still needs to understand the feedback sound and decide where they can go ahead by himself. Thus, such systems are hard to ensure the blind making a right decision according to the feedback sound. Focusing on the above problem, three kinds of auditory cues, which is converted from the traversable direction (produced by the multi-sensor fusion based obstacle avoiding algorithm) were developed in this paper for directly guiding the user where to go.

Since the weak sighted people have some degree of visual perception, and vision can provide more information than other senses, e.g. touch and hearing, the visual enhancement, which uses the popular AR (Augmented Reality) [10], [11] technique for displaying the surroundings and the feasible direction on the eyeglasses, is proposed to help the users to avoid the obstacle.

The rest of the paper is organized as follows. Section II reviews the related works involved in guiding the visually impaired people. The proposed smart guiding glasses are presented in Section III. Section IV shows some experimental results, and demonstrates the effectiveness and robustness of the proposed system. Finally, some conclusions are drawn in Section V.

This work was supported by the CloudMinds Technologies Inc.
Jinqiang Bai is with Beihang University, Beijing, 10083, China (e-mail: baijinqiang@buaa.edu.cn).
Shiguo Lian, Zhaoxiang Liu, Kai Wang are all with AI Department, CloudMinds Technologies Inc., Beijing, 100102, China (e-mail: { scott.lian, robin.liu, kai.wang }@cloudminds.com).
Dijun Liu is with DT-LinkTech Inc., Beijing, 10083, China (e-mail: liudijun@datang.com).

## II. Related Work

As this work focuses on the obstacle avoidance and the guiding information feedback, the related work with respect to such two fields are reviewed in this section.

### A. Obstacle Avoidance

There exist a vast literature on obstacle detection and avoidance. According to the sensor type, the obstacle avoidance method can be categorized as: ultrasonic sensor based method [12], laser scanner based method [13], and camera based method [5], [6], [14]. Ultrasonic sensor based method can measure the distance of obstacle and compare it with the given distance threshold for deciding whether to go ahead, but it cannot determine the exact direction of going forward, and may suffer from interference problems with the sensors themselves if ultrasonic radar (ultrasonic sensor array) is used, or other signals in indoor environment. Although laser scanner based method is widely used in mobile robot navigation for their high precision and resolution, the laser scanner is expensive, heavy, and with high power consumption, so it is not suitable for wearable navigation system. As for camera based method, there are many methods based on different cameras, such as mono-camera, stereo-camera, and RGB-D camera. Based on the mono-camera, some methods process RGB image to detect obstacles by e.g., floor segmentation [15], [16], deformable grid based obstacle detection [8], etc. However, these methods cost so much computation that they are not satisfied for real-time applications, and hard to measure the distance of the obstacle. To measure the distance, some stereo-camera based methods are proposed. For example, the method [17] uses local window based matching algorithms for estimating the distance of obstacles, and the method [18] uses genetic algorithm to generate dense disparity maps that can also estimate the distance of obstacles. However, these methods will fail under low-texture or low-light scenarios, which cannot ensure the secure navigation. Recently, RGB-D cameras have been widely used in many applications [5], [14], [19]-[21] for their low cost, good miniaturization and ability of providing wealthy information. The RGB-D cameras provide both dense range information from active sensing and color information from passive sensor such as standard camera. The RGB-D camera based method [5] combines range information with color information to extend the floor segmentation to the entire scene for detecting the obstacles in detail. The one in [14] builds a 3D (3 Dimensional) voxel map of the environment and analyzes 3D traversability for obstacle avoidance. But these methods are constrained to non-transparent objects scenarios due to the imperfection of the depth camera.

### B. Guiding Information Feedback

There are three main techniques for providing guiding information to visually impaired people [22], i.e., haptic, audio and visual. Haptic feedback based systems often use vibrators on a belt [7], helmet [23] or in a backpack [14]. Although they have far less interference with sensing the environment, they are hard to represent complicated information and require more training and concentration. Audio feedback based systems utilize acoustic patterns [8], [9], semantic speech [24], different intensities sound [25] or spatially localized auditory cues [26]. The method in [8], [9] directly maps the processed RGB image to acoustic patterns for helping the blind to perceive the surroundings. The method in [24] maps the depth image to semantic speech for telling the blind some information about the obstacles. The method in [25] maps the depth image to different intensities sound for representing obstacles in different distance. The method in [26] maps the depth image to spatially localized auditory cues for expressing the 3D information of the surroundings. However, the user will misunderstand these auditory cues under noisy or complicated environment. Visual feedback based systems can be used for the partially sighted individuals due to its ability of providing more detailed information than haptic or audio feedback based systems. The method in [27] maps the distance of the obstacle to brightness on LED (Light Emitting Diode) display as a visual enhancement method to help the users more easily to notice the obstacle. But, the LED display only shows the large obstacle due to its low resolution.

In this paper, a novel multi-sensor fusion based obstacle avoiding algorithm is proposed to overcome the above limitations, which utilizes both the depth sensor and ultrasonic sensor to find the optimal traversable direction. The output traversable direction is then converted to three kinds of auditory cues in order to select an optimal one under different scenarios, and integrated in the AR technique based visual enhancement for guiding the visually impaired people.

## III. The Proposed Framework

### A. The Hardware System

The proposed system includes a depth camera for acquiring the depth information of the surroundings, an ultrasonic rangefinder consisting of an ultrasonic sensor and a MCU (Microprogrammed Control Unit) for measuring the obstacle distance, an embedded CPU (Central Processing Unit) board acting as main processing module, which does such operations as depth image processing, data fusion, AR rendering, guiding sound synthesis, etc., a pair of AR glasses to display the visual enhancement information and an earphone to play the guiding sound. The hardware configuration of the proposed system is illustrated in Fig. 1, and the initial prototype of the system is shown in Fig. 2.

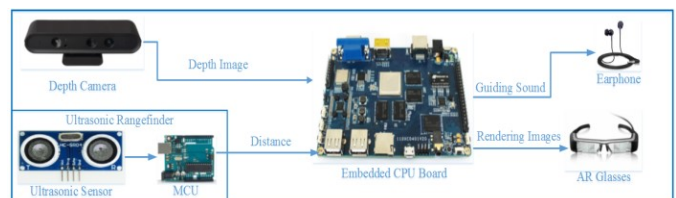

Fig. 1. The hardware configuration of the proposed system.

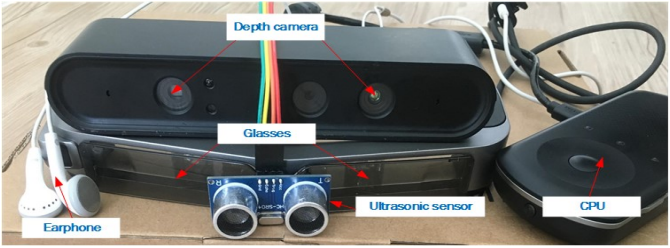

Fig. 2. The initial prototype of the proposed smart guiding glasses.

*1) Depth Information Acquisition*

Depth information is acquired with the depth sensor (the initial prototype only uses the depth camera of RGB-D camera, which includes a depth sensor and an RGB camera). The depth sensor is composed by an infrared laser source that project non-visible light with a coded pattern combined with a monochromatic CMOS (Complementary Metal Oxide Semiconductor) image sensor that captures the reflected light. The algorithm that deciphers the reflected light coding generates the depth information representing the scene. In this work, the depth information is acquired by mounting the depth sensor onto the glasses with an approximate inclination of 30°, as shown in Fig. 3. This way, considering the height of the camera to the ground to be about 1.65 m and the depth camera working range to be limited about from 0.4 m to 4 m, the valid distance in field of view is about 2.692 m, starting about 0.952 m in front of the user.

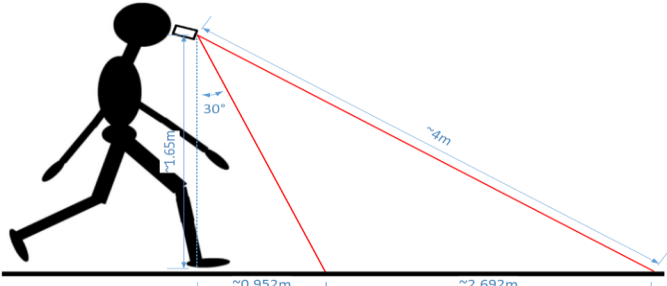

Fig. 3. Depth information acquisition.

*2) Ultrasonic Rangefinder*

In this work, the ultrasonic sensor is mounted on the glasses. The sensor uses 40 KHz samples. The samples are sent by the transmitter of the sensor. The object reflects the ultrasound wave and the receiver of the sensor receives the reflected wave. The distance of the object can be obtained according to the time interval between the wave sending and the receiving. As is shown in Fig. 4., the Trig pin of the sensor must receive a pulse of high (5 V) for at least 10 us to start measurement that will trigger the sensor to transmit 8 cycles of ultrasonic burst at 40 KHz and wait for the reflected burst. When the sensor has sent the 8 cycles burst, the Echo pin of the sensor is set to high. Once the reflected burst is received, the Echo pin will be set to low, which produces a pulse at the Echo pin. If no reflected burst is received within 30ms, the Echo pin stays high. Thus, the distance will be set very large for representing that there is no object in front of the user. The MCU is used to control the ultrasonic sensor to start measurement and detect the pulse at the Echo pin. The width of the pulse at the Echo pin is proportional to the time interval, from which the distance of the object is determined:

$$d = \frac{v \cdot ToF}{2}, \qquad (1)$$

where $d$ is the distance of the object, $v$ is the speed of sound in air, usually taken as 340 m/s and $ToF$ is the time interval between the Trig and Echo transitions.

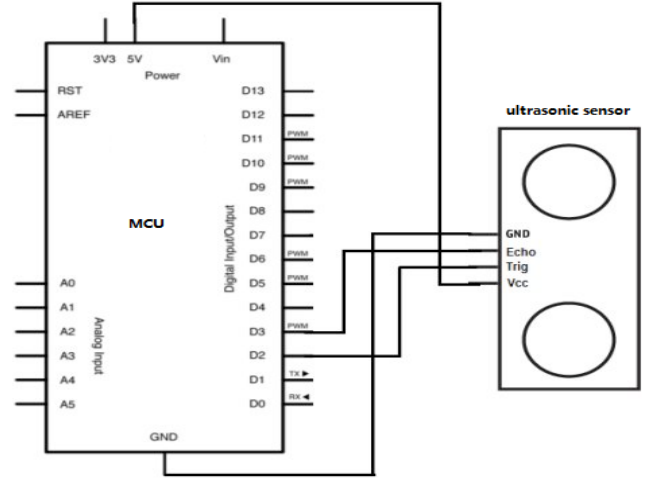

Fig. 4. Connections of ultrasonic rangefinder.

### B. The Steps

The overall algorithm diagram is depicted in Fig. 5. The depth image acquired from the depth camera is processed by the depth-based way-finding algorithm which outputs several candidate moving directions. The multi-sensor fusion based obstacle avoiding algorithm then uses the ultrasonic measurement data to select an optimal moving direction from the candidates. The AR rendering utilizes one depth image to generate and render the binocular images as well as the moving direction to guide the user efficiently. The guiding sound synthesis takes the moving direction as the input to produce the auditory cue for guiding the totally blind people. Three kinds of auditory cues are developed and tested to allow the selection of the most suitable one under different scenarios.

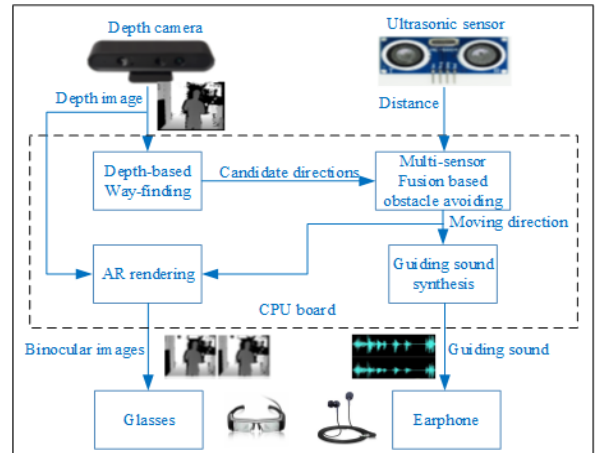

Fig. 5. Diagram of the proposed system.

*1) Depth-based Way-finding*

This depth-based way-finding algorithm is to find candidate traversable directions based on the depth image. Different from the floor-segmentation based way-finding methods, it only uses the region of interest to determine the traversable directions. Since the nearest obstacle is always at the bottom of the depth image, it only select a line in the bottom of the image as input, as is shown in Fig. 6. Considering that the user's walking is slow and gradual, it can detect the obstacle timely.

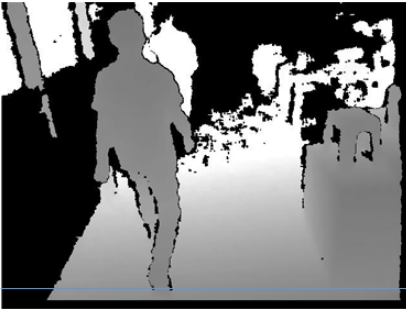

Fig. 6. The used depth image. The blue line represents the input of the depth-based way-finding algorithm.

Since the depth is relative to the camera, i.e. it is in the camera coordinate system. As is shown in Fig. 7, $O_c$ is the origin of the camera coordinate system $(X_c, Y_c, Z_c)$, i.e. the center of projection. $O$ is the origin of the image coordinate system $(u,v)$ in pixel. $O_I(u_0, v_0)$ is the principal point, i.e. the origin of the image coordinate system $(x, y)$ in millimeter. The distance from $O_c$ to the image plane is the focal length $f$. A 3D point in camera coordinates $N(x_1, y_1, z)$ is mapped to the image plane $I$ at the intersection $n(u_1, v_1)$ of the ray

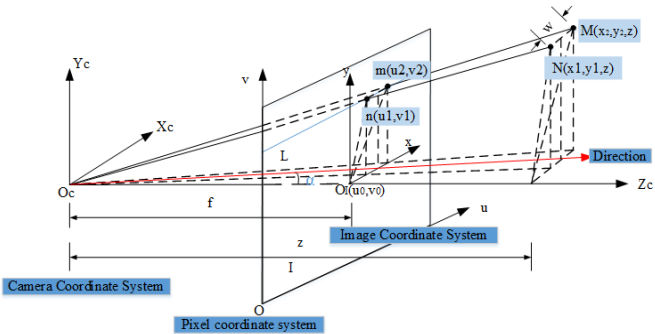

Fig. 7. Coordinate system transformation.

connecting the 3D point $N$ with the center of projection $O_c$.

The depth-based way-finding algorithm uses the traversable threshold $w$ and adaptive sliding window to determine the candidate moving directions. The sliding window size is $1 \times D(z)$, which $D(z)$ represents the adaptive width depending on the depth $z$. Every sliding step is computed as follows.

First, compute the corresponding 3D point of a given point in the depth image. As is shown in Fig. 7, given a point $n$ in the depth image, the $u_1, v_1, z$ can be known. Using similar triangles law, the 3D point $N(x_1, y_1, z)$ can be calculated by:

$$\begin{pmatrix} x_1 \\ y_1 \\ z \end{pmatrix} = \frac{z}{f} \begin{pmatrix} u_1 - u_0 \\ v_1 - v_0 \\ f \end{pmatrix}. \qquad (2)$$

Second, compute the sliding window width $D(z)$ in the image. According to the traversable threshold $w$, the 3D boundary point $M(x_2, y_2, z)$ of the traversable region can be obtained by:

$$\begin{pmatrix} x_2 \\ y_2 \\ z \end{pmatrix} = \begin{pmatrix} x_1 + w \\ y_1 \\ z \end{pmatrix}. \qquad (3)$$

Using similar triangles law as well, the 2D point $m$ of the 3D point M projection on the depth image can be computed by:

$$\begin{pmatrix} u_2 \\ v_2 \\ 1 \end{pmatrix} = \frac{f}{z} \begin{pmatrix} x_2 \\ y_2 \\ z/f \end{pmatrix} + \begin{pmatrix} u_0 \\ v_0 \\ 0 \end{pmatrix}. \qquad (4)$$

Substituting (2), (3) into (4), we can obtain:

$$\begin{pmatrix} u_2 \\ v_2 \end{pmatrix} = \begin{pmatrix} u_1 \\ v_1 \end{pmatrix} + \begin{pmatrix} fw/z \\ 0 \end{pmatrix}. \qquad (5)$$

Then the width $D(z)$ of adaptive sliding window can be expressed as:

$$D(z) = u_2 - u_1 = \frac{fw}{z}. \qquad (6)$$

Third, judge if the region between point $n$ and $m$ in the depth image is traversable. This can be calculated by:

$$1_x(z) = \begin{cases} 1 & if \forall x_{0:4} \in \{x \mid z_x \in [z_x - \varepsilon, z_x + \varepsilon], z_x > \delta\}; \\ 0 & others. \end{cases} \qquad (7)$$

where $x$ is the point in the depth image between point $n$ and $m$, $x_{0:4}$ represents continuous five points, $z_x$ is the depth of the point $x$, $\varepsilon$ is the measurement noise and set as fixed value, $\delta$ is the distance threshold.

If arbitrary continuous five points between point $n$ and $m$ is in the range $[z - \varepsilon, z + \varepsilon]$, and the depths of the five points exceed the distance threshold $\delta$ for timely and safely avoiding the obstacle, this region is considered to be traversable; otherwise, an obstacle is considered in this region, and this region will be discarded.

Fourth, compute the steering angle $\alpha$. If $1_x(z)$ in (7) is 1, i.e. the region is traversable, the steering angle $\alpha$ can be calculated by:

$$\alpha = \arctan \frac{u_1 + u_2 - 2u_0}{2f}. \qquad (8)$$

If $1_x(z)$ in (7) is 0, i.e. the region is not traversable, the steering angle $\alpha$ is not calculated.

These four steps are continually conducted until all the input points were traversed. Then the candidate direction set $A(\alpha)$, i.e. the set of steering angle $\alpha$, will be stored for later use.

*2) Multi-sensor Fusion Based Obstacle Avoiding*

Because the depth camera projects the infrared laser for measuring the distance, and the infrared laser can pass through transparent objects, which will produce incorrect measuring data. Thus, the multi-sensor fusion based method, which utilizes the depth camera and the ultrasonic sensor, is proposed and can overcome the above limitation of depth camera. This algorithm steps are as follows.

Firstly, compute the optimal moving direction based on the depth image. The optimal moving direction can be obtained by minimizing the cost function, which is defined as:

$$\alpha_{opt} = \begin{cases} \min_{\alpha \in A(\alpha)} f(\alpha) = \min_{\alpha \in A(\alpha)} (\lambda|\alpha| + \mu \frac{1}{W(\alpha)}) & if A(\alpha) \neq \varnothing; \\ Null & if A(\alpha) = \varnothing. \end{cases} \quad (9)$$

where $\alpha_{opt}$ is the optimal moving direction, $\alpha$ is the steering angle, which belongs to the $A(\alpha)$ (see section III.B.(1)), $W(\alpha)$ is the maximum traversable width which centers on the direction $\alpha$, $\lambda, \mu$ are the different weights.

The function $f(\alpha)$ evaluates the cost of both steering angle and traversable region width. The smaller the steering angle is, the faster the user can turn. The wider the traversable region is, the safer it will be. This cost function will ensure the user move effectively and safely.

Second, fuse ultrasonic data to determine the final moving direction. Since the ultrasonic sensor can detect the obstacles in the range of 0.03 m to 4.25 m, and within scanning field of 15°, the final moving direction is defined as:

$$\alpha'_{opt} = \begin{cases} \alpha_{opt} & if (\alpha_{opt} \notin [-7.5, 7.5]) \| \quad -7.5, 7.5] \&\& d > \delta); \\ Null & others. \end{cases} \quad (10)$$

where $\alpha'_{opt}$ is the final moving direction, $\alpha_{opt}$ is equal as (9), $d$ is the distance measured by the ultrasonic sensor, $\delta$ is the same as (7).

This can be explained as follows. First, it judges if the optimal moving direction in (9) is within the view field of ultrasonic sensor, i.e. [-7.5°, 7.5°]. If false, it will directly output the optimal moving direction as in (9). If it is true, the ultrasonic data then will be used to judge if the measurement distance exceeds the distance threshold $\delta$. If true, it will also output the optimal moving direction as in (9). If false, it will output Null, which means no moving direction. The workflow of this algorithm is shown in Fig. 8.

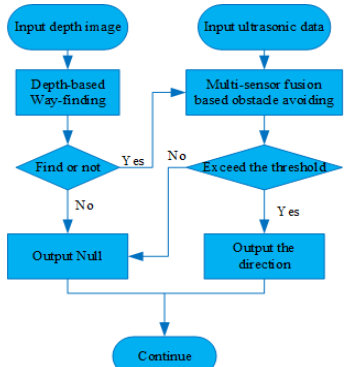

Fig. 8. The workflow of the proposed algorithm.

The results of the optimal moving direction is shown in Fig. 9, and show the multi-sensor fusion based method can make a correct decision under transparent scenario, whereas the method only by depth image cannot.

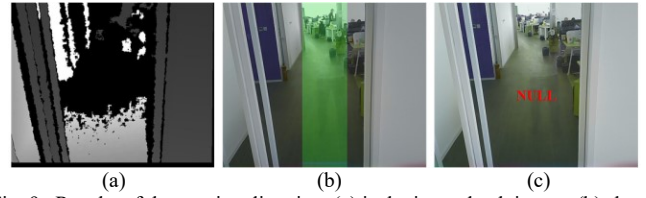

(a)                    (b)                    (c)
Fig. 9. Results of the moving direction. (a) is the input depth iamge. (b) shows the moving direction (the laurel-green region in the image) calculated in (9). (c) shows the moving direciton (Null) calculated in (10).

*3) AR Rendering with Guiding Cue*

The visual enhancement, which adopts the AR technique, is used for weak sighted people. In order to showing the guiding cue to the user based on the one depth image, the binocular parallax images are needed to generate. This was realized in Unity3D [28] by adjusting the texture coordinates of the depth image. The rendering stereo images integrate the feasible direction (the circle in Fig. 10(a) (b)) for guiding the user. When the feasible direction is located in the bounding box (the rectangular box in Fig. 10 (a) (b)), the user can go forward (see (c) of the third row in Fig. 10). When the direction is out of the bounding box, the user should turn left (see the second row in Fig. 10) or right (see the last row in Fig. 10) according to the feasible direction until it is lay in the bounding box. When the feasible direction is absent (see the first row in Fig. 10), this indicates there is no traversable way in the field of view, the user should stop and turn left or right slowly, even turn back in order to find a traversable direction.

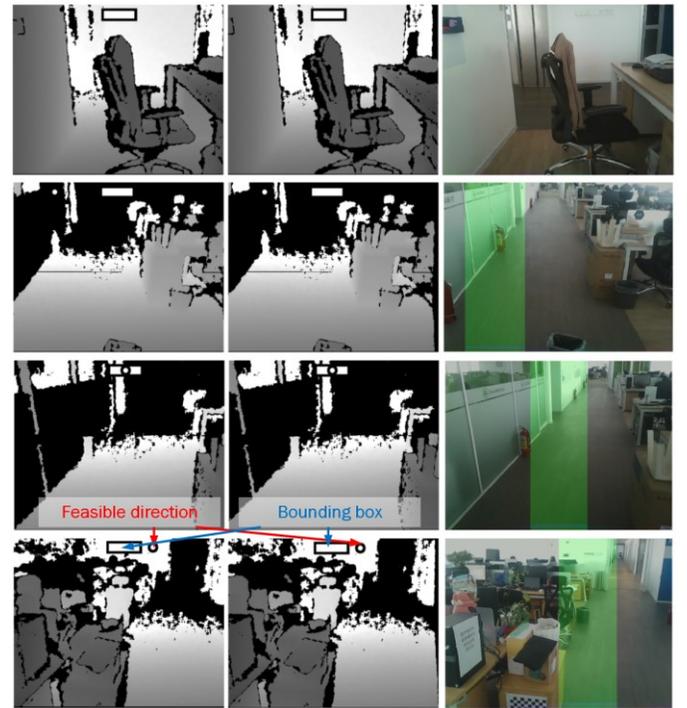

(a) Left image      (b) Right image      (c) RGB image
Fig. 10. The rendering images. (a)(b) are used for display, (c) is just for intuitive representation.

*4) Guiding Sound Synthesis*

For the totally blind users, auditory cues are adopted in this work. The guiding sound synthesis module can produce three kinds of guiding signal: stereo tone [26], recorded instructions and different frequency beep.

First kind converts the feasible direction into stereo tone. The stereo sound (see loudspeaker in Fig. 11) is like a person in the right direction to tell the user came to him. Second kind uses the recorded speech to tell the user turn left or right, or go forward. As is shown in Fig. 11, the field of view is 60°, the middle region is 15° and the two sides are divided equally. When an obstacle is in front of the user, the recorded speech will tell the user "Attention, obstacle in front of you, turn left 20 degrees". Some recorded audio instructions are detailed in TABLE I. The last one converts the feasible direction into different frequency beep. The beep frequency is proportional to the steering angle. When the user should turn left, the left channel of the earphone will work and the right will not, vice versa. When the user should go forward, the beep sound will keep silence.

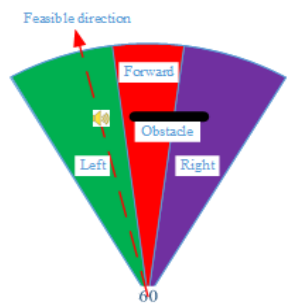

Fig .11 Guiding sketch

These three kinds of guiding method are tested in next section. Tested results show that they have their virtues and their faults. Different guiding method can be adopted in different scenarios. Next section will describe in detail.

TABLE I
AUDIO INSTRUCTIONS

| Condition | Audio instruction |
|---|---|
| Obstacle placed in front of the user with no feasible direction | Attention, obstacle in front of you, turn left or right slowly |
| Obstacle placed in front of the user with feasible direction on the left | Attention, obstacle in front of you, turn left xx[a] degrees |
| Obstacle placed in front of the user with feasible direction on the right | Attention, obstacle in front of you, turn right xx[a] degrees |
| Obstacle placed in left of the user with feasible direction on the front | Attention, obstacle in left of you, go straight |
| Obstacle placed in right of the user with feasible direction on the front | Attention, obstacle in right of you, go straight |
| Obstacle placed in left of the user with feasible direction on the right | Attention, obstacle in left of you, turn right xx[a] degrees |
| Obstacle placed in right of the user with feasible direction on the left | Attention, obstacle in right of you, turn left xx[a] degrees |
| No obstacle | Go straight |

[a] xx is the steering angle.

## IV. EXPERIMENTAL RESULTS AND DISCUSSIONS

The performance of the proposed system have been evaluated both objectively and subjectively. For the objective tests, the adaptability, the correctness, the computational cost of the proposed algorithm were analyzed. For subjective tests, 20 users (10 of them are amblyopia and the other 10 are totally blind) whose heights range from 1.5 m to 1.8 m were invited to attend the study in three main scenarios, i.e. home, office and supermarket. The main purpose of these subjective tests is to check the efficiency of the guiding instructions.

### A. Adaptability for Different Height

To test the adaptability for different user's height, we set an obstacle from 1 m to 2 m in front of the depth camera. Then the minimum detectable height of the obstacle is measured under different camera height from 1.4 m to 1.8m. The results are shown in Fig. 12, and indicate that the proposed algorithm can detect the obstacle whose height is more than 5 cm. When the obstacle's distance to the camera or the user is fixed, the lower the height of the camera (i.e. the user's height), the smaller obstacle which can be detected will be. When the height of the camera is fixed, the closer the obstacle's distance to the camera, the smaller the obstacle which can be detected will be. This is due to the size of the obstacle in the depth image is affected by the camera's height and the distance between the obstacle and the camera. Since the proposed algorithm is based on the depth image, the obstacle's size in the depth image can affect the correctness of the proposed algorithm. As is shown in Fig. 3 and Fig. 6, the distance of the region of interest is about 1-1.3 m to the user, so the minimum detectable height of the obstacle is 3 cm (see Fig. 12). Because very few object is less than 3 cm in the three main scenarios (home, office, supermarket), the proposed algorithm has very high adaptability.

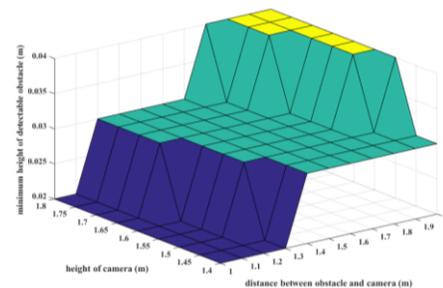

Fig. 12. Minimum detectable height of the obstacle under different height and distance

### B. Correctness of Obstacle Avoiding Algorithm

In order to evaluate the correctness of the proposed algorithm, especially under the transparent obstacle, several transparent scenarios (see Fig. 13) are selected as the test environment. Two groups of experiments were conducted, including the avoiding algorithms with and without the ultrasonic sensor. The results (see Fig. 14) reveal that the avoiding algorithm without ultrasonic sensor has the accuracy of 98.93% under the frosted glass, but has very low accuracy when encountering the pure transparent glass. This is due to the limitation of the depth camera as explained in section III.B.(4). The algorithm with the ultrasonic sensor can improve the accuracy significantly. Thus it verifies that the proposed algorithm can detect the obstacles robustly and avoid the obstacles accurately.

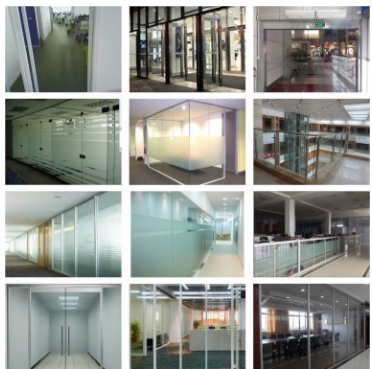

Fig. 13 Examples of different transparent obstacles

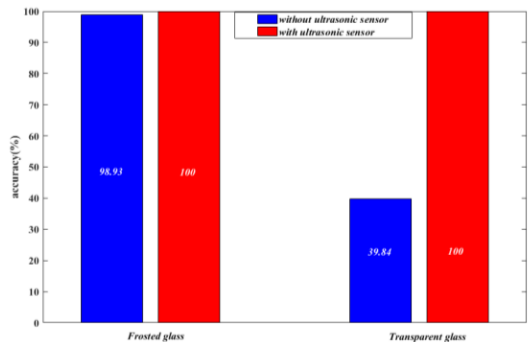

Fig. 14 Accuracy under different transparent glass

## C. Computational Cost

The average computational time for each step of the proposed system is calculated, and the results are shown in TABLE II. The depth image acquisition and the depth based way-finding algorithm takes about 11 ms. The ultrasonic sensor measurement cost depends on the obstacle's distance, which maximally takes about 26.5 ms. The ultrasonic sensor fusion algorithm takes about 1.33 ms. The AR rendering takes about 2.19 ms. Because the ultrasonic sensor measurement runs on the MCU, the multi-sensor fusion based obstacle avoiding algorithm is parallel with the ultrasonic sensor measurement, the maximum cost for processing each frame is about 30.2 ms. Since the computation can be finished in real time, the obstacle can be detected timely and the user's safety can be guaranteed sufficiently.

TABLE II
COMPUTATIONAL TIME FOR EACH STEP OF THE PROPOSED ALGORITHM

| Processing Step | Average Time |
| --- | --- |
| Depth image acquisition | 8.23 ms |
| Way-finding | 2.71 ms |
| Ultrasonic sensor measurement | Max 26.5 ms |
| Multi-sensor Fusion | 1.33 ms |
| AR rendering | 2.19 ms |
| Depth image acquisition | 8.23 ms |
| Way-finding | 2.71 ms |

## D. Interactive Experience

To test the interactive experience of the proposed system, three kinds of guiding instructions for the totally blind users were compared under three main scenarios. The experiments with and without vision enhancement proposed in this work were conducted by the weak sighted users, under the same three main scenarios (see Fig. 15). The total length of path in the home (see Fig. 15 (a)) is 40 m, and 10 kinds of obstacles (whose height is from 5 cm to 1 m) are placed on the path for testing the obstacle avoiding algorithm. The length of the path in the office (see Fig. 15 (b)) is in total 150m and the length of the path in the supermarket (see Fig. 15 (c)) amounts to 1 km. 15 kinds of obstacles are placed on the path both in the office and the supermarket.

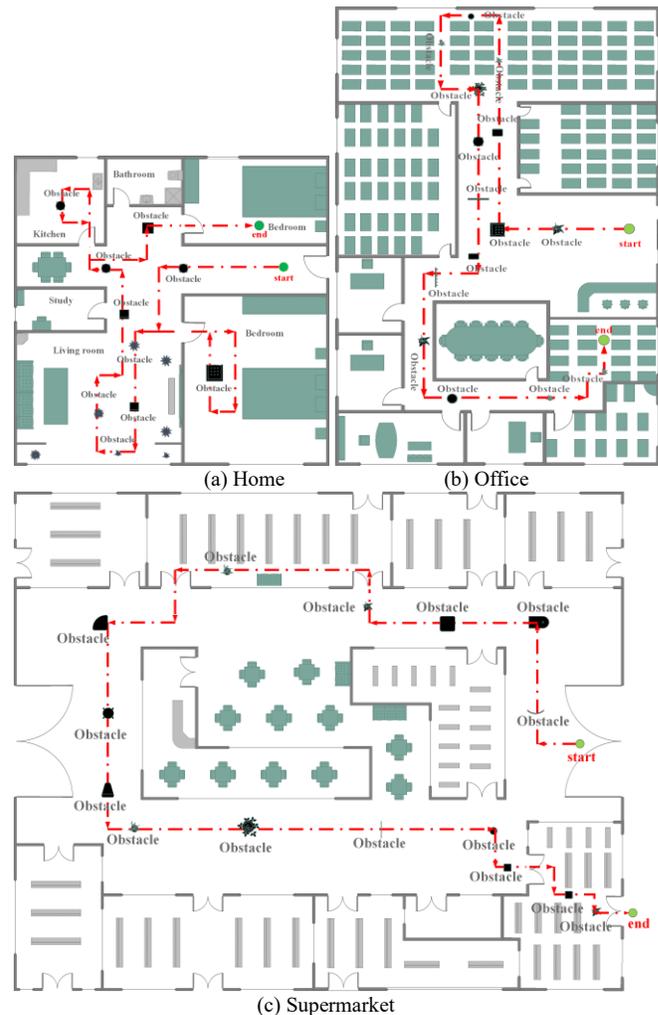

(a) Home   (b) Office

(c) Supermarket

Fig. 15 Test path under different scenarios. The red dot line is the waking path.

First, the totally blind persons with the smart guiding glasses were asked to walk in the three scenarios under three kinds of guiding instructions (see section III.B.(4)). Then the totally blind persons with a cane instead of the smart guiding glasses repeated in the three scenarios. The walking time under these scenarios was recorded respectively and are shown in TABLE III. It can be seen that when the user is in the home and office, the time cost with the stereo tone and beep sound is almost the same as that with the cane. The stereo tone based guiding instructions are more efficient than the recorded instructions based one, the beep sound based one is the most efficient.

According to the user's experience, they feel that it is hard to turn the accurate angle, therefore the recorded instructions based guiding method is not efficient enough. The cane based method is a little more efficient than the stereo tone based and recorded instructions based method, and this is because the users are familiar with their home and office. Although they do not know the obstacles on the path, they can soon make a decision with previous memory about the environment. However when they are in the unfamiliar environment, such as the supermarket, the proposed method in this work is much more efficient than the cane based one. This is because the proposed method can directly inform the user where they should go, but the cane based method must sweep the road for detecting and avoiding the obstacles, which is time-consuming. Interestingly, the recorded instructions based method is more efficient than the stereo tone based method in the supermarket. On the basis of the user's experience, this is because the supermarket is noisy relative to the home and office, the user can not identify the direction according to the stereo tone, but the recorded instructions based method can directly tell the user turn left or right. Above all, the beep sound based method is more efficient and has a better adaptability.

TABLE III
COMPUTATIONAL TIME FOR EACH STEP OF THE PROPOSED ALGORITHM

| Scenarios | Smart Guiding Glasses | | | Cane |
|---|---|---|---|---|
| | Stereo Tone | Recorded Instructions | Beep Sound | |
| Home | 91.23 s | 100.46 s | 90.08 s | 90.55 s |
| Office | 312.79 s | 350.61 s | 308.14 s | 313.38 s |
| Supermarket | 2157.50 s | 2120.78 s | 2080.91 s | 2204.15 s |

The test of visual enhancement for the weak sighted users is similar with test for the totally blind except that the guiding cues are obtained by the AR rendering images instead of the audio. Besides, the weak sighted users without the smart guiding glasses or the cane were also tested as a contrast. The results are shown in TABLE IV. From the results, we can see that when the user is in the home or the office, the time costs with the smart guiding glasses and with nothing are almost equal. But the total collisions numbers of using nothing are much more than the numbers of using the smart guiding glasses. This is because they are familiar with their home and office, the time costs can be almost the same. But the small obstacles on the ground are very hard to observe for them without the smart guiding glasses, therefor they suffer collisions more frequently. When they are in the supermarket, the time costs and the total collisions with the smart guiding glasses are much less than the one with nothing. This is because they are unfamiliar with the supermarket and have difficulty in watching the small obstacles.

TABLE IV
AVERAGE WALKING TIME AND TOTAL COLLISIONS IN DIFFERENT SCENARIOS FOR WEAK SIGHT USERS

| Scenarios | Smart Guiding Glasses | | None | |
|---|---|---|---|---|
| | Time Costs | Total Collisions | Time Costs | Total Collisions |
| Home | 74.66 s | 0 | 73.81 s | 12 |
| Office | 280.02 s | 0 | 284.57 s | 28 |
| Supermarket | 1890.50 s | 0 | 2004.03 s | 20 |

Both the totally blind and weak sighted persons' experiments verified that the proposed smart guiding glasses is very efficient and security, and very helpful for the visually impaired people in the complicated indoor environment.

V. CONCLUSION

This paper presents a smart guiding device for visually impaired users, which can help them move safely and efficiently in complicated indoor environment. The depth image and the multi-sensor fusion based algorithms solve the problems of small and transparent obstacle avoiding. Three main auditory cues for the totally blind users were developed and tested in different scenarios, and results show that the beep sound based guiding instructions are the most efficient and well-adapted. For weak sighted users, visual enhancement based on AR technique was adopted to integrate the traversable direction into the binocular images and it helps the users to walk more quickly and safely. The computation is fast enough for the detection and display of obstacles. Experimental results show that the proposed smart guiding glasses can improve the travelling experience of the visually impaired people. The sensors used in this system are simple and with low cost, making it possible to be widely used in consumer market.

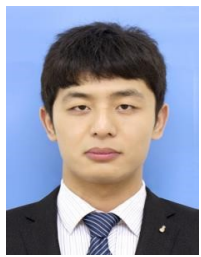
**Jinqiang Bai** got his B.E. degree and M.S. degree from China Uiversity of Petroleum in 2012 and 2015, respectively. He has been a Ph.D. student in Beihang University since 2015. His research interests include computer vision, deep learning, robotics, AI, etc.

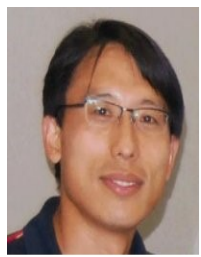
**Shiguo Lian** got his Ph.D. from Nanjing University of Science and Technology, China. He was a research assistant in City University of Hong Kong in 2004. From 2005 to 2010, he was a Research Scientist with France Telecom R&D Beijing. He was a Senior Research Scientist and Technical Director with Huawei Central Research Institute from 2010 to 2016. Since 2016, he has been a Senior Director with CloudMinds Technologies Inc. He is the author of more than 80 refereed international journal papers covering topics of artificial intelligence, multimedia communication, and human computer interface. He authored and co-edited more than 10 books, and held more than 50 patents. He is on the editor board of several refereed international journals.

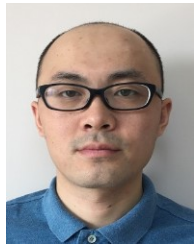
**Zhaoxiang Liu** received his B.S. degree and Ph.D. degree from the College of Information and Electrical Engineering, China Agricultural University in 2006 and 2011, respectively. He joined VIA Technologies, Inc. in 2011. From 2012 to 2016, he was a senior researcher in the Central Research Institute of Huawei Technologies, China. He has been a senior engineer in CloudMinds Technologies Inc. since 2016. His research interests include computer vision, deep learning, robotics, and human computer interaction and so on.

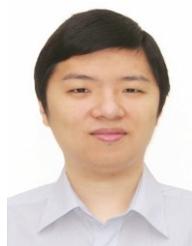
**Kai Wang** has been a senior engineer in CloudMinds Technologies Inc. since 2016. Prior to that, he was with the Huawei Central Research Institute. He received his Ph.D. degree from Nanyang Technological University, Singapore in 2013. His research interests include Augmented Reality, Computer Graphics, Human-Computer Interaction and so on. He has published more than ten papers on international journals and conferences.

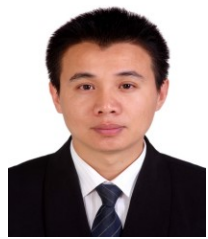
**Dijun Liu** has been a chief scientist and engineer in Datang Telecom since 2008. He was a Ph.D supervisor in Beihang University, received many awards for scientific and technological advancement. His research interests include IC Design, Image Processing, AI, Deep Learning, UAV and so on.